\documentclass[prd,preprint,nofootinbib]{revtex4}
\usepackage{amssymb}
\usepackage{epsfig}
\newcommand{\nc}{\newcommand}
\nc{\renc}{\renewcommand}
\usepackage{calc}
\usepackage{ifthen}
\usepackage{graphicx}
\usepackage{epsfig}
\usepackage{wrapfig}
\usepackage{subfigure}
\usepackage{rotfloat}
\usepackage{epsfig}
\usepackage{graphicx}
\usepackage{epsf}
\nc{\half}{{\textstyle{1\over2}}}
\nc{\etal}{\mbox{\it et al. }}
\nc{\ie}{{\it i.e.}}
\nc{\eg}{{\it e.g.}}
\renc{\thefootnote}{\arabic{footnote}}
\nc{\capt}[1]{{\bf Figure.} {\small\sl #1}}
\nc{\eqs}[2]{\mbox{Eqs.~(\ref{#1},\,\ref{#2})}}
\nc{\eq}[1]{\mbox{Eq.~(\ref{#1})}}
\nc{\figs}[2]{\mbox{Figs.~(\ref{#1},\,\ref{#2})}}
\nc{\fig}[1]{\mbox{Fig~.(\ref{#1})}}

\nc{\tag}[1]{\label{#1} \marginpar{{\footnotesize #1}}}
\nc{\mtag}[1]{\label{#1} \mbox{\marginpar{{\footnotesize #1}}}}
\renc{\baselinestretch}{1.5}
\newlength{\overeqskip}
\newlength{\undereqskip}
\nc{\be}[1]{\begin{equation} \mbox{$\label{#1}$}}
\nc{\bea}[1]{\begin{eqnarray} \mbox{$\label{#1}$}}
\nc{\Section}[2]{\section{#2}\label{#1}}
\nc{\Bibitem}[1]{\bibitem{#1}}
\nc{\Label}[1]{\label{#1}}
\nc{\eea}{\vspace{\undereqskip}\end{eqnarray}}
\nc{\ee}{\vspace{\undereqskip}\end{equation}}
\nc{\bdm}{\begin{displaymath}}
\nc{\edm}{\end{displaymath}}
\nc{\dpsty}{\displaystyle}
\nc{\bc}{\begin{center}}
\nc{\ec}{\end{center}}
\nc{\ba}{\begin{array}}
\nc{\ea}{\end{array}}
\nc{\bab}{\begin{abstract}}
\nc{\eab}{\end{abstract}}
\nc{\btab}{\begin{tabular}}
\nc{\etab}{\end{tabular}}
\nc{\bit}{\begin{itemize}}
\nc{\eit}{\end{itemize}}
\nc{\ben}{\begin{enumerate}}
\nc{\een}{\end{enumerate}}
\nc{\bfig}{\begin{figure}}
\nc{\efig}{\end{figure}}
\nc{\arreq}{&\!=\!&}
\nc{\arrmi}{&\!-\!&}
\nc{\arrpl}{&\!+\!&}
\nc{\arrap}{&\!\!\!\approx\!\!\!&}
\nc{\non}{\nonumber\\*}
\nc{\align}{\!\!\!\!\!\!\!\!&&}
\def\lsim{\; \raise0.3ex\hbox{$<$\kern-0.75em
      \raise-1.1ex\hbox{$\sim$}}\; }
\def\gsim{\; \raise0.3ex\hbox{$>$\kern-0.75em
      \raise-1.1ex\hbox{$\sim$}}\; }
\nc{\DOT}{\hspace{-0.08in}{\bf .}\hspace{0.1in}}
\nc{\Laada}{\hbox {$\sqcap$ \kern -1em $\sqcup$}}
\nc\loota{{\scriptstyle\sqcap\kern-0.55em\hbox{$\scriptstyle\sqcup$}}}
\nc\Loota{{\sqcap\kern-0.65em\hbox{$\sqcup$}}}
\nc\laada{\Loota}
\nc{\qed}{\hskip 3em \hbox{\BOX} \vskip 2ex}

\nc{\real}{{\rm I \! R}}
\nc{\Z}{{\sf Z \!\!\! Z}}
\nc{\complex}{{\rm C\!\!\! {\sf I}\,\,}}
\def\bigid{\leavevmode\hbox{\small1\kern-3.8pt\normalsize1}}
\def\id{\leavevmode\hbox{\small1\kern-3.3pt\normalsize1}}
\nc{\slask}{\!\!\!/}
\nc{\bis}{{\prime\prime}}
\nc{\pa}{\partial}
\nc{\na}{\nabla}
\nc{\ra}{\rangle}
\nc{\la}{\langle}
\nc{\goto}{\rightarrow}
\nc{\swap}{\leftrightarrow}
\nc{\EE}[1]{ \mbox{$\cdot10^{#1}$} }
\nc{\abs}[1]{\left|#1\right|}
\nc{\at}[2]{\left.#1\right|_{#2}}
\nc{\norm}[1]{\|#1\|}
\nc{\abscut}[2]{\Abs{#1}_{\scriptscriptstyle#2}}
\nc{\vek}[1]{{\rm\bf #1}}
\nc{\integral}[2]{\int\limits_{#1}^{#2}}
\nc{\inv}[1]{\frac{1}{#1}}
\nc{\dd}[2]{{{\partial #1}\over{\partial #2}}}
\nc{\ddd}[2]{{{{\partial}^2 #1}\over{\partial {#2}^2}}}
\nc{\dddd}[3]{{{{\partial}^2 #1}\over
        {\partial #2 \partial #3}}}
\nc{\dder}[2]{{{d #1}\over{d #2}}}
\nc{\ddder}[2]{{{d^2 #1}\over{d {#2}^2}}}
\nc{\dddder}[3]{{d^2 #1}\over
        {d #2 d #3}}
\nc{\dx}[1]{d\,^{#1}x}
\nc{\dy}[1]{d\,^{#1}y}
\nc{\dz}[1]{d\,^{#1}z}
\nc{\dl}[1]{\frac{d\,^{#1}l}{(2\pi)^{#1}}}
\nc{\dk}[1]{\frac{d\,^{#1}k}{(2\pi)^{#1}}}
\nc{\dq}[1]{\frac{d\,^{#1}q}{(2\pi)^{#1}}}
\nc{\cc}{\mbox{$c.c.$ }}
\nc{\hc}{\mbox{$h.c.$ }}
\nc{\cf}{cf.\ }
\nc{\erfc}{{\rm erfc}}
\nc{\Tr}{{\rm Tr\,}}
\nc{\tr}{{\rm tr\,}}
\nc{\pol}{{\rm pol}}
\nc{\sign}{{\rm sign}}
\nc{\bfT}{{\bf T }}

\def\GeV{{\rm\ GeV}}

\def\TeV{{\rm\ TeV}}
\nc{\cA}{{\cal A}}
\nc{\cB}{{\cal B}}
\nc{\cD}{{\cal D}}
\nc{\cE}{{\cal E}}
\nc{\cG}{{\cal G}}
\nc{\cH}{{\cal H}}
\nc{\cL}{{\cal L}}
\nc{\cO}{{\cal O}}
\nc{\cT}{{\cal T}}
\nc{\cN}{{\cal N}}
\nc{\rvac}[1]{|{\cal O}#1\rangle}
\nc{\lvac}[1]{\langle{\cal O}#1|}
\nc{\rvacb}[1]{|{\cal O}_\beta #1\rangle}
\nc{\lvacb}[1]{\langle{\cal O}_\beta #1 |}
\nc{\bb}{\bar{\beta}}
\nc{\bt}{\tilde{\beta}}
\nc{\ctH}{\tilde{\cal H}}
\nc{\chH}{\hat{\cal H}}
\nc{\1}{\aa}
\nc{\2}{\"{a}}
\nc{\3}{\"{o}}
\nc{\4}{\AA}
\nc{\5}{\"{A}}
\nc{\6}{\"{O}}
\nc{\al}{\alpha}
\nc{\g}{\gamma}
\nc{\Del}{\Delta}
\nc{\e}{\epsilon}
\nc{\eps}{\epsilon}
\nc{\lam}{\lambda}
\nc{\om}{\omega}
\nc{\Om}{\Omega}
\nc{\ve}{\varepsilon}
\nc{\mn}{{\mu\nu}}
\nc{\vp}{\varphi}
\nc{\rf}[1]{(\ref{#1})}
\nc{\nn}{\nonumber \\*}
\nc{\bfB}{\bf{B}}
\nc{\bfv}{\bf{v}}
\nc{\bfx}{\bf{x}}
\nc{\bfy}{\bf{y}}
\nc{\vx}{\vec{x}}
\nc{\vy}{\vec{y}}
\nc{\oB}{\overline{B}}
\nc{\oI}{\overline{I}}
\nc{\oR}{\overline{R}}
\nc{\rar}{\rightarrow}
\nc{\ti}{\times}
\nc{\slsh}{\hskip-5pt/}
\nc{\sm}{Standard~Model~}
\nc{\MP}{M_{\rm Pl}}
\nc{\tp}{t_{\rm Pl}}
\nc{\ave}{\bar{E}}

\nc{\eff}{{\rm eff}}
\nc{\kk}{\vek{k}}
\nc{\pp}{{\rm p}}
\nc{\ga}{g_{a\gamma}}
\nc{\vv}{\\}
\nc{\eee}{{\bf E}}
\nc{\bbb}{{\bf B}}
\nc{\qcd}{T_{\rm QCD}}
\nc{\G}{\rm \ G}
\def\vec#1{{\bf #1}}
\def\lae{\;^{<}_{\sim} \;} \def\gae{\; ^{>}_{\sim} \;} 

\def\ell{e^{c}LL}

\begin{document}
\title{ 
Supersymmetric inflation and baryogenesis via Extra-Flat directions of the MSSM
}
\author{John McDonald}
\email{j.mcdonald@lancaster.ac.uk}
\affiliation{ 
Cosmology and Astroparticle Physics Group,
University of Lancaster,
Lancaster LA1 4YB, United Kingdom
}
\author{Osamu Seto}
\email{osamu.seto@uam.es}
\affiliation{
 Instituto de F\'{i}sica Te\'{o}rica UAM/CSIC, 
 Universidad Aut\'{o}noma de Madrid, Cantoblanco, Madrid 28049, Spain
}
%
\begin{abstract}
   One interpretation of proton stability is that it implies the existence of extra-flat directions of the minimal supersymmetric standard model, in particular $u^{c}u^{c}d^{c}e^{c}$ and $QQQL$, where the operators lifting the potential are suppressed by a mass scale $\Lambda$ which is much larger than the Planck mass, $ \Lambda \gae 10^{26} \GeV$. Using $D$-term hybrid inflation as an example, we show that such flat directions can serve as the inflaton in supersymmetric inflation models. The resulting model is a minimal version of $D$-term inflation which requires the smallest number of additional fields. In the case where $Q$-balls form from the extra-flat direction condensate after inflation, successful Affleck-Dine baryogenesis is possible if the suppression mass scale is $\gae 10^{31}-10^{35} \GeV$. In this case the reheating temperature from $Q$-ball decay is in the range $3-100 \GeV$, while observable baryon isocurvature perturbations and non-thermal dark matter are possible. In the case of extra-flat directions with a large $t$ squark component, there is no $Q$-ball formation and reheating is via conventional condensate decay. In this case the reheating temperature is in the range $1-100 \TeV$, naturally evading thermal gravitino overproduction while allowing sphaleron erasure of any large $B-L$ asymmetry.    
\end{abstract}
\pacs{}
\preprint{IFT-UAM/CSIC-08-04} 
\vspace*{1.5cm}
\maketitle
\section{Introduction}

Successful models of supersymmetric (SUSY) inflation should ideally satisfy
 a number of requirements: natural compatibility 
with supergravity (SUGRA), lack of fine-tuned couplings, 
successful post-inflation era including reheating and baryogenesis, and compatibility with 
unified models of particle physics. With respect to these conditions, SUSY hybrid inflation models have
a particular attraction \cite{dti,fti}. They can achieve sufficient inflation without requiring very small or fine-tuned couplings, and in the case of $D$-term hybrid inflation they are naturally compatible with SUGRA \cite{eta}. 
   Focusing on the $D$-term inflation case, a natural question is the origin of the fields in the $D$-term inflation sector. The $U(1)$ gauge field and charged vector pair $\Phi_{\pm}$ of $D$-term inflation might be understood as components of an extended gauge theory. However, the inflaton is usually a gauge singlet which is added to the model for no other reason\footnote{Models exist which attempt to identify the inflaton with a known field, such as a right-handed sneutrino \cite{mury,antusch,kadota}.}. If we do not add such a singlet, can $D$-term hybrid inflation still occur? Here we argue that it can. The vector pair will naturally couple to any gauge-invariant combination of fields in the MSSM. Such gauge-invariant products (monomials) also characterise flat directions of the MSSM. Thus a natural possibility is that a flat direction can play the role of the inflaton in $D$-term inflation models\footnote{An interesting model using MSSM flat directions as inflatons, which has a quite different philosophy with respect to fine-tunings, is given in \cite{AEGM:AtermInf}. See also \cite{lytha}.}. In this model the number of additional fields required for inflation is reduced to just a $U(1)$ gauge field and the $\Phi_{\pm}$ vector pair, so providing a minimal version of $D$-term inflation.  
    As we will show, conventional MSSM flat directions lifted by Planck scale-suppressed gauge-invariant superpotential terms are unsuitable. This is because such terms generally lift the flat direction scalar at field strengths well below the value required for inflation. However, it is known that certain gauge-invariant superpotential terms must be suppressed by more than the Planck scale or forbidden entirely. The $d = 4$ operators $u^{c}u^{c}d^{c}e^{c}$ and $QQQL$ will lead to rapid proton decay if they are only Planck scale-suppressed \cite{protond}. 
 One way this problem can be solved is by assuming that the underlying complete theory introduces a dynamical suppression factor into the non-renormalisable superpotential interactions, such that the effective mass scale suppressing the dangerous operators is $ \Lambda \gae 10^{26} \GeV$ \cite{protond}. 
It is also possible that this dynamical suppression will also apply to all higher-order MSSM superpotential terms lifting the flat direction, such as  $(u^{c}u^{c}d^{c}e^{c})^{2}$ and $(QQQL)^{2}$. We will refer to a flat direction for which this is true as an `extra-flat direction'. An alternative interpretation of the absence of proton decay is in terms of a discrete symmetry which eliminates the dangerous $d = 4$ operators \cite{luhn}. In this case it is possible that the higher-order operators will be unsuppressed. However, as we will show, such unsuppressed flat directions, even if higher-order, cannot serve as an inflaton.    
  If the existence of extra-flat directions is the correct interpretation of the absence of proton decay in the MSSM, then an extra-flat direction scalar could serve as the inflaton in a $D$-term inflation model. The 
extra-flat direction potential at large field values is naturally lifted to an inflationary plateau by its gauge-invariant superpotential coupling to $\Phi_{+}\Phi_{-}$. Reheating and possibly baryogenesis would then come from the decay of the flat direction inflaton, via either $Q$-ball decay or conventional homogeneous condensate decay, depending on the $t$ squark component of the flat direction. 
In this paper we will study $D$-term inflation along an extra-flat direction of the MSSM. The paper is organised as follows. In Section 2 we discuss extra-flat directions and the resulting $D$-term inflation model. In Section 3 we discuss reheating and baryogenesis. In Section 4 we present our conclusions. 

\section{$D$-term inflaton along extra-flat directions}
\subsection{Potential}
  We consider a flat direction $\Phi$ in the MSSM and
 introduce two additional fields $\Phi_{\pm}$ charged 
 under a U(1) gauge group with the Fayet-Illiopoulos term $\xi$.
The superpotential is
\be{superpotential}
W = \frac{\lambda_{1} \Phi^m}{m M^{m-3}}+ \frac{\lambda_{2} \Phi^n}{n M^{n-1}}\Phi_+ \Phi_-    ~,
\ee
where $\lambda_{1,2}$ are Yukawa couplings and $M$ is the reduced Planck mass, $M = M_{Pl}/\sqrt{8 \pi}$~\footnote{
A SUSY mass term $W \supset \mu \Phi_+\Phi_-$ has not been included. This term would induce $n$ minima with 
nonvanishing VEV for $\Phi$, which consist of squark and/or slepton VEV and lead to large baryon or lepton number violation in the MSSM. Although in most cases there is no symmetry which can exclude such a term, we note that for the case $m = n$ this term can be excluded by an R-symmetry which allows the terms $\Phi^{n}$ and $\Phi^{n}\Phi_{+}\Phi_{-}$. In addition, if $\mu$ is less than the scale of soft SUSY breaking terms, the minimum of the potential can be at $\Phi = 0$, while for larger $\mu$ there can be directions in the complex $\Phi$ plane along which the field evolution can avoid the minima with $\Phi \neq 0$.}. 
We will present results for general $m$ and $n$, specialising to the case of most interest $m = n = 4$, corresponding to $\Phi^4 \sim u^{c}u^{c}d^{c}e^{c}$ or $QQQL$. Proton stability in the case of  $u^{c}u^{c}d^{c}e^{c}$ or $QQQL$ requires that $\lambda_{1} \lae 10^{-8}$, corresponding to an effective suppression mass scale $\Lambda = M/\lambda_{1} \gae 10^{26} \GeV$. However, $\lambda_{2}$ is unconstrained by phenomenology and will be determined by the inflation model.  The scalar potential in the global SUSY limit is then
\begin{eqnarray}
V &=& \left|  \frac{\lambda_{2} \phi^n}{n M^{n-1}} \right|^2(|\phi_+|^2+|\phi_-|^2)
  + \left| \frac{\lambda_{1} \phi^{m-1}}{M^{m-3}}  +  \frac{ \lambda_{2}\phi^{n-1}}{M^{n-1}}\phi_+\phi_- \right|^2 \nonumber \\
 && + \frac{g^2}{2}(\xi+|\phi_+|^2-|\phi_-|^2)^2 .
\label{GlobalScalarPotential}
\end{eqnarray}
The supersymmetric global minimum is located at\footnote{Note that there is a SUSY flat direction when $\mu = 0$ and $\phi = 0$, such that $|\phi_{+}|^{2} - |\phi_{-}|^{2} = \xi$. However, the minimum with $\phi_{+} = 0$ is selected since $\phi_{+}$ gains a large mass when $\phi \neq 0$ during inflation.} 
\begin{eqnarray}
 (\phi, \phi_+, |\phi_-|)  =  ( 0, 0, \sqrt{\xi} ).
\end{eqnarray}
If 
\begin{equation}
 \left|\frac{\lambda_{1} \lambda_{2}  \phi^{m+n-2}}{M^{m+n-4}}\right|^2 \ll (g^2 \xi)^2 
\label{NegligibleMixing}
\end{equation}
 is satisfied, the mixing between $\phi_+$ and $\phi_-$ is negligible.
The potential is then simplified to
\begin{eqnarray}
V  \simeq  \left|  \frac{\lambda_{2} \phi^n}{n M^{n-1}} \right|^2(|\phi_+|^2+|\phi_-|^2) + \left| \frac{ \lambda_{1} \phi^{m-1}}{M^{m-3}} \right|^2 
  + \frac{g^2}{2}(\xi+|\phi_+|^2-|\phi_-|^2)^2 .
\label{SimpleGlobalPotential}
\end{eqnarray}
The critical value of $\phi$ is given by
\begin{equation}
 |\phi_{\rm c}| \equiv \left( \frac{n M^{n-1} \sqrt{g^2 \xi} }{|\lambda_{2}|}\right)^{1/n}   ~,
\label{phi_c:Definition}
\end{equation} 
 which determines the stability of the $\phi_-$ field at the origin. 
The origin is a false vacuum for $|\phi| > |\phi_c|$, while 
 it is unstable for $|\phi| < |\phi_{\rm c}|$. 

\subsection{Inflationary expansion}
For $|\phi| > |\phi_{\rm c}|$, $\phi_- =0$ is a local minimum
 and there is the false vacuum energy from the $D$-term, which drives inflation.
The potential during inflation is given as
\begin{equation}
 V \simeq \frac{1}{2}g^2 \xi^2 \left(1 + \frac{g^2}{8 \pi^2}
 \ln\frac{\sigma^{2n}}{\Lambda_{*}^{2n}} \right)
\label{SimplestInflationPotential}
~,\end{equation}
 where $\sigma = \sqrt{2}Re(\phi)$ is the canonically normalised inflaton
 and $\Lambda_{*}$ is the renormalisation scale.
Inflation ends when the inflaton reaches the larger
 of $\sigma_{\rm c} \equiv \sqrt{2}|\phi_{rm c}|$ and
\begin{equation}
 \sigma_f \equiv \frac{\sqrt{n} g M}{2 \pi} ~,
\end{equation}
where $\sigma_{f}$ corresponds to the end of slow-roll.
However, a non-vanishing F-term potential is also present in this model.
Hence, we need to ensure that the condition $V_F \ll V_D$ is satisfied, which requires that
\begin{equation}
 \left|\frac{\lambda_{1} \phi^{m-1}}{M^{m-3}}\right|^2 \ll \frac{1}{2} g^2 \xi^2
\label{DtermDomination}
\end{equation}
 is satisfied. Note that when this is satisfied, equation (\ref{NegligibleMixing}) is also satisfied.
The dynamics of the inflaton field is similar to that in 
the minimal $D$-term hybrid inflation model~\cite{dti}. 
The solution of the slow-roll field equations is
\be{z1}  
\sigma^{2}(N) = \sigma_0^2 + \frac{n g^2 N M^2}{2 \pi^2}     ~.
\ee
Here, $\sigma_0 = {\rm max}[\sigma_f, \sigma_{\rm c}]$ is the expectation value of inflaton when inflation terminates.
The spectral index is
\be{z2} 
n_{s} = 1 - \frac{1}{N}
 \left(1 + \frac{2\pi^2 \sigma_0^2}{n g^2 M^2 N} \right)^{-1}      ~,
\ee
while the value of $\xi^{1/2}$ normalised to the curvature perturbation 
$P_{\zeta}$ is 
\be{z3} 
\frac{\xi^{1/2}}{M} =   \left(\frac{3 n P_{\zeta}}{N}\right)^{1/4}   
\left(1 + \frac{2\pi^2 \sigma_0^2}{n g^2 M^2 N} \right)^{-1/4}     ~.
\ee
In the case of $\sigma_0 = \sigma_{\rm c}$, which we will show is true in examples of interest, we find
\begin{equation}
\frac{2\pi^2 \sigma_0^2}{n g^2 M^2 N} = \left( \frac{\lambda_{2_{c}}}{\lambda_{2}} \right)^{2/n} 
\end{equation}
with
\be{z4} 
\lambda_{2_{\rm c}} = \left( \frac{4 \pi^2}{n g^2 N M^2} \right)^{n/2}  
\left( g^2 \xi n^2 M^{2 \left(n-1\right)} \right)^{1/2}      ~.
\ee    
$\lambda_{2} = \lambda_{2_{c}}$ corresponds to $\sigma_{N} = 2 \sigma_{c}$, 
with $\sigma \approx \sigma_{c}$ throughout inflation 
when $\lambda_{2} < \lambda_{2_{c}}$.
When $\lambda_{2} \gg \lambda_{2_{c}}$ as well as the case of $\sigma_0 = \sigma_{\rm c}$, the spectral index is $n_{s} = 1- 1/N \approx 0.98$, as in conventional $D$-term inflation, while the value of $\xi^{1/2}$
required to account for the observed curvature perturbation ($P_{\zeta}^{1/2} =
4.8 \times 10^{-5}$) is  
$ \xi^{1/2} =   7.9 \times 10^{15} n^{1/4} \GeV $.
On the other hand, in the case where $\lambda_{2} \ll \lambda_{2_{c}}$, the spectral index approaches $n_{s} = 1$ while the value of $\xi^{1/2}$ is reduced by a factor $(\lambda_{2}/\lambda_{2_{c}})^{1/2n}$.

\subsection{Comparison with observations}  

The spectral index observed by WMAP, $n_{s} = 0.958 \pm 0.016$ (1-$\sigma$) \cite{wmap},  is substantially smaller the $D$-term inflation value. In addition, WMAP data permits at most an O(10)$\%$ contribution to the CMB power spectrum from cosmic strings \cite{csbound,csbound2,csbound3}, which implies that $\xi^{1/2} \lae 4 \times 10^{15} \GeV$. (Here we have used $G \mu = 2 \times 10^{-6}$ for the $l = 10$ WMAP normalised string tension \cite{bevis}.) One way to interpret the WMAP observations is that they correspond to an adiabatic curvature perturbation with $n_{s} \approx 1$ combined with a 10$\%$ cosmic string contribution \cite{csinterp}, which can be achieved by making $\lambda_{2}$ sufficiently small compared with $\lambda_{2_{c}}$. In this case the apparent spectral index of the combined perturbation is effectively lowered and can be in agreement with the 3-year WMAP data analysis \cite{csinterp}. It is a striking feature of $D$-term inflation models in general that they have a solution which increases $n_{s}$ while decreasing the cosmic string contribution, just as required for this interpretation of the WMAP observations.   
With respect to this possibility, the extra-flat direction model has a possible advantage over conventional $D$-term inflation. The contribution of cosmic strings to the CMB power spectrum is proportional to $\mu^2 = (2 \pi \xi)^2$. In the case of conventional $D$-term inflation with $n_{s} = 1$, the value of $\xi^2$ in the limit $\lambda_{2} \ll \lambda_{2_{c}}$ is proportional to $\lambda_{2}^2$. Therefore $\lambda_{2}$ must lie within a rather narrow range of values for the cosmic string contribution to be O(10)$\%$. In the case of the extra-flat direction inflaton, the dependence is $ \propto \lambda_{2}^{2/n}$. Therefore the CMB contribution varies much more gradually with $\lambda_{2}$ e.g.  
$\xi^2 \propto \lambda_{2}^{1/2}$ for the case $n = 4$. Thus an O(10)$\%$ contribution is obtained for a much wider range of $\lambda_{2}$, making it perhaps a more natural possibility than in conventional $D$-term inflation.

\subsection{Constraints from cosmic string bound, SUGRA and potential flatness}

     We first check that the 10$\%$ cosmic string condition 
$\xi^{1/2} \approx 4 \times 10^{15} \GeV$ can be satisfied for
reasonable values of $g$ when $|\phi_{c}|^2$ is small enough compared 
with $M^2$ for SUGRA corrections to be neglected.
We will require that  $|\phi_{c}| <  k M$, with $k \lae 0.3$, so that $|\phi_{c}|^{2} \lae 0.1 M^2$. From equation~(\ref{phi_c:Definition}), this implies that
\be{a1}
 g \lae \frac{\lambda_{2} k^{n} M}{n \xi^{1/2}}        ~.
\ee
For the case $n = 4$ and $\xi^{1/2} \approx 4  \times 10^{15} \GeV$, 
equation~(\ref{a1}) implies that
\be{a2}
 g \lae 1.2 \lambda_{2} \left(\frac{k}{0.3}\right)^{4}     ~.
\ee
Thus $\lambda_{2}$ should not be small compared with 1 if $g$ is not very small compared with 1.  
From equation (\ref{z3}), to suppress $\xi^{1/2}$ from $7.9 \times 10^{15} \GeV$ to $4 \times 10^{15} \GeV$ in the case $n = 4$ we require that $\lambda_{2_{c}}/\lambda_{2} \approx  250$, which implies that
\be{a3} 
\lambda_{2} \approx 8 \times 10^{-7} g^{-3}    ~.
\ee
Equations (\ref{a2}) and (\ref{a3}) imply that
\be{a4}
 g \lae 0.03 \left(\frac{k}{0.3}\right) ~
\ee
and
\be{a44}
 \lambda_{2} \gae 0.03  \left(\frac{0.3}{k}\right)^{3}   ~.
\ee
Thus $k \gae 0.1$ is necessary when $\lambda_{2} \lae 1$ in order to satisfy equation~(\ref{a44}). $0.1 \lae k \lae 0.3$ then implies that $g \approx 0.01-0.03$ and $0.03 \lae \lambda_{2} \lae 1$.    
Therefore, as in conventional $D$-term inflation in the small coupling limit, $g$ must be somewhat smaller than the Standard Model gauge couplings \cite{csbound2}. In addition, $\lambda_{2}$ must be much larger than $\lambda_{1}$~\footnote{We have assumed that $\sigma_{c} > \sigma_{f}$. For the case $n = 4$ this requires that $\xi^{1/2}/M > |\lambda_{2}|g^{3}/\pi^{4}$.
With $\xi^{1/2} \approx 4 \times 10^{15} \GeV$ and $g \approx 0.02$ this is easily satisfied.}. 
     We next evaluate Eq.~(\ref{DtermDomination}) to find the condition 
on $\lambda_{1}$ for F-term corrections not to spoil the flatness of 
the inflaton potential. In general we find
\be{a4a} 
\lambda_{1} \ll \frac{1}{\sqrt{2}} \frac{g \xi}{k^{m-1} M^2}     ~.
\ee
For the case $m = 4$ this gives
\be{a4b} 
\lambda_{1} \ll 7 \times 10^{-5} g \left(\frac{0.3}{k}\right)^{3} \left(\frac{\xi^{1/2}}{4 \times 10^{15} \GeV}\right)^{2}      ~.
\ee
Thus for values of $\lambda_{1}$ which satisfy the proton decay constraint, 
$\lambda_{1} \lae 10^{-8}$,
the flat direction potential is easily sufficiently flat to serve as an inflaton. However, for the case of unsuppressed $n = m = 4$ flat directions with $\lambda_{1} \sim 1$, the F-term would violate the flatness of the flat-direction inflaton potential.
             An alternative solution of the proton decay problem is to consider elimination of the $m = n = 4$ operators
entirely by a symmetry. In this case we expect to have unsuppressed operators with $m = n = 8$, such that $\lambda_{1} \sim \lambda_{2} \sim 1$. However, in this case the F-flatness condition will still be violated. For $n = 8$, equation~(\ref{a1}) implies that
\be{p1} 
g \lae 5 \times 10^{-3} \lambda_{2} \left( \frac{k}{0.3} \right)^{8}       ~.
\ee
To suppress $\xi^{1/2}$ to $4 \times 10^{15} \GeV$,
with $n = 8$ we need $\lambda_{2_{c}}/\lambda_{2} \approx 6.5 \times 10^4$. This implies that
\be{p2}      
\lambda_{2} = 9 \times 10^{-12} g^{-7}      ~.
\ee
Therefore if $\lambda_{2} \lae 1$ we have $g \gae 0.03$. Equation~(\ref{p2}) combined with equation~(\ref{p1}) gives
\be{p3} 
g \lae 0.02 \left( \frac{k}{0.3} \right)    ~.
\ee
Thus $k \gae 0.5$ is necessary if $g \gae 0.03$.
The F-term flatness condition equation~(\ref{a4a}) for $m = 8$ is
\be{p4} 
 \lambda_{1} \lae   8 \times 10^{-6}  \left( \frac{g}{0.03} \right)\left( \frac{0.5}{k} \right)^{7}   \left( \frac{\xi^{1/2}}{4 \times 10^{15} \GeV} \right)^{2}      ~.
\ee
 Thus the $m = n = 8$ flat direction will also need to be extra-suppressed to have a flat inflaton potential, even if the $m = n = 4$ term is completely eliminated by a discrete symmetry. Therefore extra-flat directions are essential for an MSSM flat direction to play the role of the inflaton in $D$-term inflation.

\subsection{Post-inflationary evolution}

Including soft SUSY breaking terms, the potential is
\begin{eqnarray}
V &=& m_{\phi}^2|\phi|^2 + A m_{3/2}\frac{\lambda_{1} \phi^m}{m M^{m-3}} + {\rm H.c.}
  \nonumber \\
 && + g^2 \xi |\phi_-|^2 \left|\frac{\phi}{\phi_{\rm c}} \right|^{2n}
  + \left| \frac{\lambda_{1} \phi^{m-1}}{M^{m-3}}\right|^2
  + \frac{g^2}{2}(\xi-|\phi_-|^2)^2 ,
\label{PotentialAfterInflation}
\end{eqnarray}
 where we used equation~(\ref{phi_c:Definition}). Here
$m_{\phi}^2$ is the soft SUSY breaking scalar mass.
The potential has two SUSY non-renormalisable terms:
 the third term $g^2 \xi^2 (|\phi_-|^2/\xi)|\phi^n/\phi_{\rm c}^n|^2$ and
 the fourth term $|\lambda_{1} \phi^m/M^{m-3}|^2$.
Assuming that $m-1 < n$ (since the case of most interest will be that where $m = n$),
the third term is dominant if $\phi \gtrsim \phi_*$,  where
\begin{equation}
 \phi_*^{n-m+1} \equiv \frac{\lambda_{1} \phi_{\rm c}^n}{M^{m-3} \sqrt{g^2 \xi^2}}
  \left(\frac{\xi}{|\phi_-|^2}\right)^{1/2},
\end{equation}
while the fourth term is dominant if $\phi < \phi_*$.
    An important point in what follows is that for $\phi \gae \phi_{*}$, the $A$-term will be effectively suppressed
compared with the usual case of an MSSM flat direction with potential stabilised by a non-renormalisable term. This is because the $A$-term is coming from the first term in the superpotential, equation~(\ref{superpotential}), whereas the non-renormalisable term in the scalar potential is from the second term. As a result, the baryon asymmetry generated by the Affleck-Dine mechanism \cite{ad,drt} will be suppressed relative to the MSSM flat direction case.    
Once inflation ends, the $\phi_-$ field oscillates around the minimum $\langle\phi_{-}\rangle = \sqrt{\xi}$ \footnote{In general, the minimum of the potential is at $|\phi_{-}| = \xi^{1/2}\left(1 - |\phi/\phi_{c}|^{2n}\right)^{1/2}$. This rapidly tends to 
$|\phi_{-}| = \xi^{1/2}$ as the $\phi$ oscillations are damped from $\phi_{c}$ to small amplitudes.}.
The $\phi$ field will oscillate around the origin
 dominated by either the $|\lambda_{1} \phi^{m-1}/M^{m-3}|^2$ or the
 $ |\langle \phi_-\rangle|^2 g^2 \xi | \phi^n/\phi_{\rm c}^n |^2 $ term,
 depending on the amplitude.
While the amplitude of the $\phi$ oscillation is large, the energy density of $\phi$ will decrease more rapidly than that of $\phi_{-}$ ($V \propto \phi^{d}$ implies that $\rho \propto a^{-6d/\left(d + 2\right)}$, with $d \geq 6$ for $\phi$ oscillations and $d = 2$ for $\phi_{-}$ oscillations), so the Universe initially becomes $\phi_-$ dominated.
In the following we will assume that the $\phi_{-}$ oscillations efficiently
decay into radiation. (We will comment on how our results are altered if this is not satisfied.)  
Due to the $\phi_-$ decay, the radiation produces two distinct thermal corrections to
the potential equation~(\ref{PotentialAfterInflation}).
The $\phi$ field is expected to acquire a thermal mass term
\be{th1}
 h^2 T^2 |\phi|^2~,
\ee
 with $h$ being a coupling between $\phi$ and a particle
 in the thermal bath~\cite{ThermalMass} in the case where
 the expectation value of the field is relatively small and
 the radiation temperature is high enough, and also a logarithmic term
\be{th2}
 \alpha T^4 \ln \frac{|\phi|^2}{T^2}~, 
\ee
 which appears at the two-loop level
 through the running of couplings with non-vanishing $\phi$~\cite{TwoLoop}.
Here, $\alpha$ is a constant of order of $10^{-2}$ and
 its sign can be positive or negative.
For our purpose, hereafter we consider the case that
$\alpha$ is negative.
(For a positive $\alpha$, the field oscillates around the origin
by either the mass $m_{\phi}$, the thermal mass or this two-loop effect and
simply decays into radiation.)
          The potential with the two-loop induced logarithmic potential is
\begin{eqnarray}
V(\phi) &=& m_{\phi}^2\left(1 +K \ln \frac{|\phi|^2}{\Lambda^2}\right)|\phi|^2
 + A m_{3/2}\frac{\lambda_{1} \phi^m}{m M^{m-3}} + {\rm H.c.}
  + \alpha T^4 \ln \frac{|\phi|^2}{T^2} \nonumber \\
 && +g^2\xi |\phi_-|^2 \left|\frac{\phi^n}{\phi_{\rm c}^n} \right|^2
  + \left|\frac{\lambda_{1} \phi^{m-1}}{M^{m-3}}\right|^2   ,
\label{PotentialWithThermal}
\end{eqnarray}
 where we
 include the radiative correction to $m_{\phi}$ with
 the direction dependent coefficient $K$. For MSSM flat directions which do not include a large top quark component, $K \simeq - 10^{-2}$ ~\cite{EMD:Qball, K}. However, a large top squark component can drive $K$ to positive values.
 
A negative $K$ is the source of the spatial instability
 which leads to $Q$-ball formation
 in the gravity mediated SUSY breaking model~\cite{EMD:Qball}.
For a negative $\alpha$, as shown in Ref.~\cite{Seto:Gravitino},
 the thermal mass term cannot appear
 because of a relatively large expectation value of the field.
Here, the $\phi$ field is trapped with nonvanishing value
 by the thermal logarithmic term, equation~(\ref{th2}), 
 and the non-renormalizable term,
 until the temperature decreases to a certain value.
As the temperature falls, the expectation value of $\phi$ becomes small.
When the $\phi$ becomes as small as
\be{e16}
 |\phi_{\rm os}|^2 \simeq \frac{(-\alpha) T^4}{m_{\phi}^2} ~,
\label{InitialAmplitude}
\ee
 $\phi$ starts to oscillate around the origin with the angular momentum
 in the $\phi$ space induced by $A$-term,
 which is equivalent to the charge density (baryonic and/or leptonic)
 carried by $\phi$~\cite{ad}.
Provided that the reheating by the $\phi_-$ decay is completed
 before $\phi$ starts to oscillate, we find from equation~(\ref{InitialAmplitude}) that
\begin{equation}
 \left.\frac{\rho_{\phi}}{\rho_{\rm R}}\right|_{t_{\rm os}}
  \simeq \frac{30 (-\alpha)}{\pi^2 g_*} .
\label{Phi-Rratio}
\end{equation}
Here, $g_*$ is the effective total degrees of freedom
of the relativistic species in the radiation.
Since the ratio in equation~(\ref{Phi-Rratio}) is of order of $10^{-4}$,
the $\phi$ field oscillations (or the $Q$-ball density formed from the $\phi$
condensate if $K < 0$) soon dominates the Universe.

\section{Reheating and Baryogenesis}
     Reheating in this model is from the decay of the extra-flat direction inflaton field. The reheating temperature will therefore depend on whether or not the flat direction condensate fragments into $Q$-balls, which in turn depends on the $t$ squark content of the flat direction \cite{K}.

\subsection{$Q$-ball formation and decay}

The $\phi$ field oscillates around the origin coherently to begin with,
but there is a spatial instability of its fluctuations due to the negative $K$. After inhomogeneities in the field grow,
the coherent $\phi$ fragments and, as a result, $Q$-balls are
formed~\cite{EMD:Qball, Qball}.
Here we briefly summarize properties of $Q$-balls in
gravity mediated SUSY breaking models. The radius of a $Q$-ball, $R$, is
estimated as $R^2 \simeq 2/(|K|m_{\phi}^2)$~\cite{EMD:Qball}.
By numerical calculations, it was shown that almost all the produced charge
is stored inside $Q$-balls, and that a good fit to the $Q$-ball charge is
\be{q1}
 Q \simeq \bar{\beta}\left(\frac{|\phi_{\rm{os}}|}{m_{\phi}}\right)^2
 \epsilon_Q
\ee
 with
\begin{equation}
\epsilon_Q =
\left\{
\begin{array}{ll}
\epsilon  \;\quad {\rm for} \quad \epsilon \gtrsim \epsilon_{\rm c} \\
\epsilon_{\rm c}  \quad {\rm for} \quad \epsilon < \epsilon_{\rm c}
\end{array}
\quad ,
\right.
\end{equation}
 and
\be{eps}
\epsilon \equiv \left.\frac{n_q}{n_{\phi}}\right|_{t_{\rm{os}}} \simeq
 2q|A| \left(\frac{m_{3/2}}{m_{\phi}}\right) \sin\delta \times {\rm Min} \left[ \left(\frac{\lambda_{1}}{\lambda_{1\;*}}\right), 1 \right] \left(\frac{m_{\phi}}{H_{\rm os}}\right)~, 
\ee
 where $\delta$ is the $CP$ violating phase, $\epsilon_{\rm c} \simeq 10^{-2} $  and $\bar{\beta}= 6\times 10^{-3}$ ~\cite{KK:fittingCharge}.
(For $\epsilon < \epsilon_{c}$ the condensate will fragment to pairs of oppositely charged $Q$-balls.) The last two factors in equation~(\ref{eps}) are, respectively, the suppression of the baryon asymmetry due to the effective suppression of the $A$-term relative to the non-renormalisable term once $\phi \gae \phi_{*}$ (where $\lambda_{1\;*}$ is defined below), and the enhancement due to $H_{\rm os} \ll m_{\phi}$ at the onset of $\phi$ oscillations, which allows the $B$ violating $A$-term to act over many $\phi$
oscillations before expansion diminishes the $A$-term\footnote{In the case where the $\phi_{-}$ field does not decay efficiently to radiation, the Universe after inflation will
be dominated by $\phi_{-}$ oscillations and onset of oscillations will be typically determined by an order $H^{2}$ correction to the $\phi$ mass squared due to non-minimal K\"ahler interactions of the form $|\phi_{-}|^2|\phi|^2$. In this case $m_{\phi} \approx H_{\rm os}$ in \eq{eps}.}.      
 
The decay temperature of $Q$-ball is given by~\cite{Td}
\begin{equation}
 T_{\rm d} \simeq 1 \sqrt{f_{\rm s}}\left(\frac{m_{\phi}}{1 {\rm TeV}}\right)^{1/2}
  \left(\frac{10^{20}}{Q}\right)^{1/2} {\rm GeV},
\label{DecayTemp}
\end{equation}
where $10^3 \gtrsim f_{\rm s} \geq 1$ is the enhancement factor in the decay
if $Q$-balls can decay into final states consisting purely of scalar particles.
Since $Q$-balls come to dominate the Universe in our scenario,
 the decay temperature gives the reheating temperature
 at the onset of radiation dominated Universe.
The resultant emitted charge to entropy ratio is given by
\begin{equation}
 \frac{n_q}{s} = \frac{3}{4} \frac{T_{\rm d}}{m_{\phi}}\epsilon .
\label{BE-ratio}
\end{equation}

\subsection{Affleck-Dine baryogenesis}

The baryon asymmetry is generated by the $B$ and $CP$ violating $A$-term
when $\phi$ starts to oscillate around the origin \cite{ad}.
As usual in $D$-term inflation, there is no order $H$ correction to the
$A$-term before $\phi$ starts oscillating,
since $\phi_{+} = 0$ throughout \cite{cmr}.
In addition, in the case of a non-singlet inflaton there can be
 no linear coupling of the inflaton $I$ to superpotential monomials $W$ in the K\"ahler potential of the form $I^{\dagger}W$, which would generate an order $H$ $A$-term correction \cite{drt}.  
Therefore the phase of the inflaton relative to the $A$-term at the onset of
$\phi$ oscillations, $\theta$, is determined by its initial random value during inflation, in which case $\sin \delta \approx (\sin 2 \theta)/2$. This phase gives the $CP$ violating phase required for Affleck-Dine baryogenesis, with $n_{B} \propto
\theta$ for $\theta$ small compared with 1.  
In the estimation of the resultant baryon asymmetry produced
 by the Affleck-Dine mechanism,
 the important quantity is the amplitude of the AD field
 when it starts to oscillate, $\phi_{\rm os}$.
For $\phi_{\rm os } \gae \phi_*$,
 the amplitude is given by
\begin{equation}
 m_{\phi}^2 \simeq
  n g^2 \xi^2 \left|\frac{\phi_{\rm os}^{n-1}}{\phi_{\rm c}^n}\right|^2 ~,
\end{equation}
 where we assume $|\phi_-|^2=\xi$.
On the other hand, for $\phi_{\rm os } \lae \phi_*$,
 the amplitude is given by
\begin{equation}
 m_{\phi}^2 \simeq (m-1) \left|\frac{\lambda_{1} \phi_{\rm os}^{m-2}}{M^{m-3}}\right|^2.
\end{equation}
The former applies in the case of a small $\lambda_{1} \lae \lambda_{1\;*}$, with
\begin{equation}
\lambda_{1\;*}^{(n-1)} \equiv
\left(\frac{\sqrt{g^2\xi^2}}{|\phi_{\rm c}|^n}\right)^{m-2}
\left(\frac{m_{\phi}^2}{n}\right)^{\frac{n-m+1}{2}}
M^{\left(m-3\right)\left(n-1\right)} ,
\end{equation}
 while the latter corresponds to a large $\lambda_{1} \gae \lambda_{1\;*}$.
 For the case $\lambda_{1} \lae \lambda_{1\;*}$ we obtain
\begin{equation}
 \frac{|\phi_{\rm os}|^2}{m_{\phi}^{2}}
  = \left(\frac{1}{n g^2 \xi^2}\frac{|\phi_{\rm c}|^{2n}}{m_{\phi}^{2(n-2)}}\right)^{\frac{1}{n-1}}.
\label{InitialAmplitude:LargeM}
\end{equation}
Equation~(\ref{q1}) then gives
\be{e26}
Q \simeq \bar{\beta} \left(\frac{1}{n g^2 \xi^2}\frac{|\phi_{\rm c}|^{2n}}{m_{\phi}^{2(n-2)}}\right)^{\frac{1}{n-1}} \epsilon_Q .
\ee

 The expansion rate at the onset of $\phi$ oscillations during radiation domination can be obtained from equations~(\ref{e16}) and (\ref{InitialAmplitude:LargeM}),
\be{h1} 
 \frac{H_{\rm os}}{m_{\phi}} = \left( \frac{\pi^{2} g_{*}}{90 \alpha} \right)^{1/2} \frac{|\phi_{\rm os}|}{M}     ~,
\ee
with
\be{h2} 
 \frac{|\phi_{\rm os}|}{M} = \left(\frac{\sqrt{n} m_{\phi}}{\lambda_{2} \xi^{1/2}}\right)^{\frac{1}{n-1}}   ~.
\ee
For $n = 4$ this gives,
\be{h3}
 \frac{H_{\rm os}}{m_{\phi}} \approx 4 \times 10^{-4} \alpha^{-1/2} \lambda_{2}^{-1/3}
\left(\frac{m_{\phi}}{1 \TeV}\right)^{1/3}
\left( \frac{4 \times 10^{15} \GeV}{\xi^{1/2}} \right)^{1/3}    ~.
\ee
Therefore $H_{\rm os}/m_{\phi} \approx 10^{-3}$. This justifies neglect of $H$ corrections to the soft SUSY breaking terms at the onset of $\phi$ oscillations.
  
For $n = 4$, equation~(\ref{e26}) becomes
\begin{equation}
 Q \simeq 1.1 \times 10^{21}
\left(\frac{0.1}{g}\right)^{2/3}
\left(\frac{4 \times 10^{15} \GeV}{\xi^{1/2}}\right)^{4/3}
  \left(\frac{|\phi_{\rm c}|}{0.3 M}\right)^{8/3}
  \left(\frac{1 {\rm TeV}}{m_{\phi}}\right)^{4/3} \epsilon_Q .
\label{Q:largeM:n=4}
\end{equation}
Then
 from equations~(\ref{DecayTemp}), (\ref{BE-ratio}) and (\ref{Q:largeM:n=4}),
and using $m_{\phi}/H_{\rm os} \approx 10^{3}$,
the decay temperature and baryon asymmetry are given by
\begin{equation}
 T_{\rm d} \simeq 3 \sqrt{f_{\rm s}} \left( \frac{0.1}{\sqrt{\epsilon_{Q}}} \right)
  \left(\frac{m_{\phi}}{1 {\rm TeV}}\right)^{7/6}
  \left(\frac{0.1}{g}\right)^{-1/3}
  \left(\frac{4 \times 10^{15} \GeV}{\xi^{1/2}}\right)^{-2/3}
  \left(\frac{|\phi_{\rm c}|}{0.3 M}\right)^{-4/3} \GeV ,
\end{equation}
and
\begin{equation}
 \frac{n_q}{s} \simeq 2 \times 10^{-10}
 \left( \frac{0.1}{\sqrt{\epsilon_{Q}}} \right)
 \left(\frac{\epsilon}{10^{-7}}\right)
  \sqrt{f_{\rm s}} \left(\frac{m_{\phi}}{1 {\rm TeV}}\right)^{1/6}
  \left(\frac{0.1}{g}\right)^{-1/3}
  \left(\frac{4 \times 10^{15} \GeV}{\xi^{1/2}}\right)^{-2/3}
  \left(\frac{|\phi_{\rm c}|}{0.3 M}\right)^{-4/3} .
\end{equation}
The observed baryon asymmetry is $n_{q}/s = \left(1.8 \pm 0.1\right) \times 10^{-10}$.
Hence $\epsilon \lae 10^{-7}$ is
necessary to account for the observed $B$ asymmetry. The $Q$-ball decay temperature,
which gives the reheating temperature, is in the range 3-100 GeV for $1 \leq f_{\rm s} \lae 10^{3}$.  
    For $\lambda_{1} \lae \lambda_{1\;*}$, from equation~(\ref{eps}) we have $\epsilon \approx (0.1-1)(\lambda_{1}/\lambda_{1\;*}) (m_{\phi}/H_{\rm os}) \theta$. The random phase of $\phi$ during inflation would be expected to be of order 1; therefore in order to generate the observed baryon asymmetry we require that $\lambda_{1} \approx (10^{-10}-10^{-9}) \lambda_{1\;*}$.
With $m=n=4$, $\lambda_{1\;*}$ is,
\begin{equation}
 \lambda_{1*} \simeq 3.3 \times 10^{-8}
 \left(\frac{m_{\phi}}{1 {\rm TeV}}\right)^{1/3}
 \left(\frac{4 \times 10^{15} \GeV}{\xi^{1/2}}\right)^{-2/3}
 \lambda_{2}^{2/3}  ~.
\end{equation}
Since $\lambda_{2}$ should not be very small if $g$ is not very small, $\lambda_{1\;*} \approx 10^{-8}$ is likely. Therefore to account for the observed baryon asymmetry with $\theta \approx 1$ we must have $\lambda_{1} \lae 10^{-17}$. This corresponds to suppression of the $QQQL$ or $u^{c}u^{c}d^{c}e^{c}$ superpotential terms by a mass scale
$\Lambda \gae 10^{35} \GeV$. Thus a much larger suppression is necessary for successful baryogenesis than is required by proton stability.
    Note that it may be possible for $Q$-balls to decay at a temperature greater than that of the electroweak transition if $f_{\rm s} \approx 10^{3}$, corresponding to $Q$-ball decay to purely scalar final states. In this case any dangerous baryon asymmetry will be erased by $B+L$ violating sphaleron fluctuations.  

\subsection{Baryon isocurvature perturbations}

    The $CP$ violating phase $\delta$ is given by the
phase of $\phi$ during inflation relative to the $A$-term, which defines the real direction.  Therefore in the case where reheating is via $Q$-ball decay, $\epsilon \approx 10^{2}(\lambda_{1}/\lambda_{1\;*}) \theta$ implies that $\theta \approx
10^{-9}(\lambda_{1\;*}/\lambda_{1})$ in order to have $\epsilon \approx 10^{-7}$, as required for successful baryogenesis with $f_{\rm s} = 1$. Quantum fluctuations of $\phi$ in the phase direction will lead to baryon isocurvature perturbations, which can be large when $\theta \ll 1$. For uncorrelated baryon isocurvature perturbations, the fractional contribution to the CMB power spectrum is given by $\alpha_{\rm BI}$, where \cite{biso,jiso}
\be{x1}
 \alpha_{\rm BI} = \left( \frac{\Omega_{B}}{\Omega_{\rm DM}} \right)^{2} \frac{f_{\theta}^2 H^2}{4 \pi^2 P_{\rm R} \phi^2}
  ~,
\ee
with $\delta n_{\rm B}/n_{\rm B} \approx f_{\theta} \delta \theta$. In our case $f_{\theta} \approx 1/\theta$. The present observational limit is $\alpha_{\rm BI} < 0.26$ \cite{bean}. With $\theta \approx 10^{-9} (\lambda_{1\;*}/\lambda_{1})$, $\Omega_{\rm DM} =0.23$ and $\Omega_{B} = 0.04$, this gives a upper bound on $H/\phi$, 
\be{x2} 
\frac{H}{2 \pi \phi} \lae 10^{-13} \left(\frac{\lambda_{1\;*}}{\lambda_{1}}\right)       ~.
\ee
$\xi^{1/2} \approx 4 \times 10^{15} \GeV$, corresponding to O(10)$\%$ 
cosmic strings, implies that $H = 2.7 \times 10^{12} g \GeV$. 
Since we are considering $\phi \approx \phi_{c} \approx (0.1-0.3) M$, 
we therefore have
\be{x3} 
 \frac{H}{2 \pi \phi} \approx (0.6 -1.8) \times 10^{-6} g      ~,
\ee
Thus with $g \approx 0.01-0.03$, the baryon isocurvature perturbation is sufficiently small if $\lambda_{1}/\lambda_{1\;*} \lae 10^{-5}$. The correct baryon asymmetry then requires that $\theta \gae 10^{-4}$. 
 Thus even if the initial random phase of the flat direction field could satisfy $\theta \ll 1$, the flat direction would still have to be suppressed by $\Lambda \gae 10^{31} \GeV$ in order to avoid large baryon isocurvature perturbations.  
For $\Lambda \approx 10^{31} \GeV$ and $\theta \approx 10^{-4}$, the correct baryon asymmetry will be generated together with a potentially observable baryon isocurvature perturbation.

\subsection{Non-thermal dark matter}

The reheating temperature is $\approx 1 \GeV$ for the case where $Q$-ball decay to purely scalar final states is kinematically suppressed, such that $f_{\rm s} = 1$. This low reheating temperature implies that $Q$-balls may decay below the freeze-out temperature of neutralino LSPs, in which case $Q$-ball decay will also produce non-thermal LSP dark matter particles. In fact dark matter particles are often overproduced,
in particular for the standard bino-like neutralino LSP.
Although several ways to avoid this problem have been proposed by taking an alternative choice of
the LSP~\cite{EDM:Singlino, FH:Higgsino, Seto:Gravitino, Seto:Axino},
perhaps the simplest ones are
to assume a Higgsino-like neutralino LSP~\cite{FH:Higgsino}, or a
gravitino LSP ~\cite{Seto:Gravitino} with
a sneutrino NLSP to escape BBN constraints~\cite{KKKM:SneuOnBBN}.

\subsection{Reheating from flat direction condensate decay without $Q$-ball formation}

    In the case where the inflaton corresponds to a flat direction with a large $t$ squark component, the $\phi$ condensate will not fragment to $Q$-balls since $K > 0$ \cite{K}. In this case reheating will occur via conventional flat direction condensate decay and a higher reheating temperature is expected. For the $B-L$ conserving $u^c u^c d^c e^c$ and $QQQL$ directions, the baryon asymmetry from Affleck-Dine baryogenesis will be erased by sphaleron $B+L$ violation so long as the $\phi$ condensate decays at $T > T_{ew}$. In this case it is possible for the initial phase of $\phi$ to take its natural value, $\theta \approx 1$, without requiring a suppression of the flat direction beyond that required to evade proton decay.
        Assuming that $\phi$ oscillations dominate the energy density when the $\phi$ field decays to radiation, the energy density is given by $\rho \approx
m_{\phi}^{2} \phi_{\rm d}^{2}$, where $\phi_{\rm d}$ is the amplitude of the oscillations when they decay.
$|\phi_{\rm d}|$ is then related to the decay temperature $T_{\rm d}$ by
\be{x4}
 |\phi_{\rm d}|^2  = \frac{k_{\rm d} T_{\rm d}^{4}}{m_{\phi}^2} \;\;\;;\;\;   
k_{\rm d} = \frac{\pi^2 g(T_{\rm d})}{30}   ~. 
\ee 
For $h |\phi_{\rm d}| > m_{\phi}$, where $h$ is the gauge or Yukawa coupling of MSSM particles to the flat direction, particles coupling to $\phi$ gain masses greater than $m_{\phi}$ and so the $\phi$ decay is kinematically suppressed. Therefore the condensate decays once $\phi \approx m_{\phi}/h$, assuming that $\Gamma_{\rm d} > H$ when this occurs. The energy density in
the field at this time is $\rho \approx m_{\phi}^{4}/h^{2}$. Therefore the decay temperature, which is equivalent to the reheating temperature $T_{R}$, is
\be{x5}
 T_{\rm d} \approx \frac{m_{\phi}}{\left(h^{2} k_{\rm d}\right)^{1/4} }        ~.
\ee  
With $g(T_{D}) \approx 200$ we find $k_{\rm d} \approx 65$. Therefore
\be{x6} T_{\rm d} \approx 1.1 \left(\frac{m_{\phi}}{1 \TeV}\right)\left(\frac{0.1}{h}\right)^{1/2} \TeV ~,
\ee
where the particles with the {\it smallest} coupling $h$ to $\phi$ will 
dominate the decay process, so long as
$\Gamma_{\rm d} > H$. Therefore $T_{\rm d} \approx 1 - 100 \TeV$ in this model, assuming that the smallest coupling satisfies $0.1 \gae h \gae 10^{-5}$. 
Once $h \phi < m_{\phi}$ the $\phi$ decay rate may be estimated 
to be $\Gamma_{\rm d} \approx h^{2} m_{\phi}/4 \pi$, so the condition
$\Gamma_{\rm d} > H \approx 5 T_{\rm d}^2/M$ is easily satisfied for $T_{\rm d}$ 
in this range.
  
    We have assumed that the kinematic suppression of 
the decay rate prevents $\phi$ decaying until $m_{\phi} \gae h \phi$, 
in which case $T_{\rm d} \lae 100 \TeV$. We should check that $\phi$ decay through 
heavy intermediate particles cannot cause it to decay significantly earlier. 
The decay rate via heavy intermediate particles of mass $h \phi$ will have 
the generic form
\be{x7} 
\Gamma_{\rm d} \approx \frac{\alpha_{\rm d} m_{\phi}^{1+ r}}{\left(h \phi\right)^{r}}     ~,
\ee
where $\alpha_{\rm d} < 1$ is a product of couplings and phase space factors.
Since there are two heavy intermediate states, $r \geq 4$ is expected. 
For $r = 4$, and using equation~(\ref{x4}), (\ref{x7}) 
and $\Gamma_{\rm d} \approx H(T_{\rm d}) \approx 5 T_{\rm d}^2/M$, 
this gives for the decay temperature
\be{x8}
 T_{\rm d} \approx \left( \frac{\alpha_{\rm d}}{g^4 k_{\rm d}^2 k_{T}} \right)^{1/10} \left(m_{\phi}^{9} M\right)^{1/10} \approx 35  \left( \frac{\alpha_{\rm d}}{g^4 k_{\rm d}^2 k_{T}} \right)^{1/10} \left( \frac{m_{\phi}}{1 \TeV} \right)^{9/10} \TeV       ~.
\ee
Thus for typical couplings, the decay through intermediate states will also result in a reheating temperature in the range $T_{\rm R} \approx 1-100 \TeV$.
     It is significant that the reheating temperature, $T_{\rm R} \lae 100 \TeV$, is naturally compatible with the thermal gravitino upper bound, $T_{R} \lae 10^{6} \GeV$, without any tuning of couplings. Even though the inflaton is part of the MSSM sector, it still leads to the required low reheating temperature. Since sphaleron $B+L$ violation will erase the baryon asymmetry produced by the flat direction inflaton decay, baryogenesis must occur via some other mechanism, such as Affleck-Dine baryogenesis along an orthogonal flat direction. 

\section{Conclusions} 

We have shown that it is possible for an MSSM extra-flat direction
 (one suppressed by an effective mass scale much larger than the Planck mass)
of the form $QQQL$ or $u^{c}u^{c}d^{c}e^{c}$ to play the role of the inflaton 
in a $D$-term inflation model. 
This eliminates the otherwise unmotivated singlet inflaton, 
reducing the number of required additional fields and so 
providing a minimal version of $D$-term inflation. 
The model has all the advantages of conventional $D$-term inflation 
with respect to compatibility with SUGRA and absence of fine-tuned couplings. 
   The nature of reheating depends on whether the extra-flat direction is unstable with respect to $Q$-ball formation. In the case where $Q$-balls form, it is possible to generate the baryon asymmetry via $Q$-ball decay so long as the mass scale suppressing the flat direction is sufficiently large, $\Lambda \gae 10^{31}-10^{35} \GeV$,  depending on the random phase $\theta$ of the flat direction scalar during inflation. With $\Lambda \approx 10^{31} \GeV$ and $\theta \approx 10^{-4}$ it is possible to generate an observably large baryon isocurvature perturbation. The reheating temperature from $Q$-ball decay is typically in the range $3-100$ GeV. As this can be less than the neutralino LSP freeze-out temperature, it is also possible to produce non-thermal dark matter from $Q$-ball decay. 
  In the case where the flat direction has a large $t$ squark component, there is no $Q$-ball formation.  In this case the reheating temperature from decay of the homogeneous flat direction condensate is in the range $1-100 \TeV$, ensuring sphaleron erasure of the baryon asymmetry from the $B-L$ conserving directions while remaining naturally compatible with the thermal gravitino upper bound on $T_{R}$. The fact that we are able to calculate the reheating temperature in this case is a direct consequence of the inflation being part of the MSSM sector. Since the baryon asymmetry from the flat direction is erased, the mass scale suppressing the flat direction in this case is constrained only by proton decay, $\Lambda \gae 10^{26} \GeV$. 
 We have interpreted the WMAP observation of the spectral index as being due to an order 10$\%$ CMB contribution from cosmic strings combined with a nearly scale-invariant adiabatic curvature perturbation, $n_{s} \approx 1$. As in conventional $D$-term inflation, we can simultaneously suppress the contribution of the cosmic strings to the required level while increasing $n_{s}$ by considering a small enough coupling of the inflaton to the Fayet-Iliopoulos charged fields. The extra-flat direction $D$-term inflation model has an advantage over conventional $D$-term inflation in that the range of coupling which leads to an order 10$\%$ contribution from cosmic strings is much wider, making it perhaps more natural. For this solution to work, it is also necessary to have a $U(1)$ gauge coupling that is somewhat smaller than the known gauge couplings, $g \approx 0.01-0.03$. A significant feature of the model is that the superpotential coupling of the monomial $QQQL$ or $u^{c}u^{c}d^{c}e^{c}$ to $\Phi_{+}\Phi_{-}$ must be much larger than the pure monomial superpotential coupling. This feature may serve to test the compatibility of the model with an ultra-violet complete theory, as we would naively expect all the superpotential couplings of the monomial to be strongly suppressed. Finally, we note that other solutions to the cosmic string and spectral index problems are possible, for example SUGRA corrections from a non-minimal K\"ahler potential \cite{osyok} and/or modification of the inflaton potential by other fields, such as a RH sneutrino \cite{lin}. 
          
   The existence of extra-flat directions of the MSSM is one way to interpret the empirical suppression of non-renormalisable MSSM superpotential terms demanded by proton stability. It will be important to establish whether extra-flat directions can be understood in the context of an ultra-violet complete theory and to explore more generally their role and possible signatures in cosmology. 
         
%
\section*{Acknowledgments}
The work of O.S. is supported by the MEC project FPA 2004-02015 
and the Comunidad de Madrid project HEPHACOS (No.~P-ESP-00346). 
The work of JM  was supported (in part) by the European Union through 
the Marie Curie Research and Training Network "UniverseNet" (MRTN-CT-2006-035863)
and by STFC (PPARC) Grant PP/D000394/1" 

\end{document}